\documentclass[useAMS,usenatbib,usegraphicx]{mn2e}
\usepackage{lscape}
\usepackage{multicol}
\usepackage{longtable}
\usepackage{pdfpages}

\usepackage{graphicx}	
\usepackage{footnote}
\usepackage{amsmath}	
\usepackage{amssymb}	
\usepackage{multicol}        
\usepackage{bm}		
\usepackage{pdflscape}	
\usepackage[T1]{fontenc}
\usepackage{ae,aecompl}

\newcommand{\RNum}[1]{\uppercase\expandafter{\romannumeral #1\relax}}

\def\hcop    {HCO$^+$}             

\def\kmss    {$km\,s^{-1}$}
\def\kms     {$km\,s^{-1}$\,}

\title[Effects of infall and outflow on massive star-forming regions.]{Effects of infall and outflow on massive star-forming regions.}

\author[Qiang Li et al.]{Qiang Li$^{1,3}$\thanks{Email: liqiang@xao.ac.cn},
Jianjun Zhou$^{1,2}$\thanks{Email: zhoujj@xao.ac.cn},
Jarken Esimbek$^{1,2}$,
Yuxin He$^{1,2}$,
Willem Baan$^{1,4}$,
Dalei Li$^{1,2}$,
\newauthor Gang Wu$^{1,2}$,
Xindi Tang$^{1,2}$,
Weiguang Ji$^{1}$,
Toktarkhan Komesh$^{1,3,5}$,
Serikbek Sailanbek$^{1,3,5}$\\
$^{1}$Xinjiang Astronomical Observatory, Chinese Academy of Sciences, Urumqi 830011, P. R. China\\
$^{2}$Key Laboratory of Radio Astronomy, Chinese Academy of Sciences, Urumqi 830011, P. R. China\\
$^{3}$University of the Chinese Academy of Sciences, Beijing 100080, P. R. China\\
$^{4}$Netherlands Institute for Radio Astronomy, 7991 PD Dwingeloo, The Netherlands\\
$^{5}$Department of Solid State Physics and Nonlinear Physics, Faculty of Physics and Technology, AL-Farabi Kazakh National University, \\Almaty 050040, Kazakhstan\\}

\date{Accepted 2019 July 23. Received 2019 June 21; in original form 2019 March 11}
\begin{document}
\maketitle

\begin{abstract}
A total of 188 high-mass outflows have been identified from a sample of 694 clumps from the Millimetre Astronomy Legacy Team 90 GHz survey, representing a detection rate of approximately 27\%. The detection rate of outflows increases from the proto-stellar stage to the H\,\RNum{2} stage, but decreases again at the photodissociation (PDR) stage suggesting that outflows are being switched off during the PDR stage. An intimate relationship is found between outflow action and the presence of masers, and water masers appear together with 6.7 GHz methanol masers. Comparing the infall detection rate of clumps with and without outflows, we find that outflow candidates have a lower infall detection rate.
Finally, we find that outflow action has some influence on the local environment and the clump itself, and this influence decreases with increasing evolutionary time as the outflow action ceases.

\end{abstract}


\section{INTRODUCTION}
Star formation is an intrinsically complex process involving the collapse and accretion of matter onto proto-stellar objects \citep{1985ARA&A..23...267} and  infall and outflow motions play an important role in the star formation processes.
However, a comprehensive understanding of both processes, particularly towards massive star-forming regions, is still lacking.
In part, this is because of the larger distances involved and the typically more clustered and complex nature of star formation regions, making it difficult to disentangle the infall and outflow properties of individual objects in a given cluster \citep{2018MNRAS.477..2455}. In this paper, we continue to study the dynamical processes in massive star-forming regions to further understand star formation processes.

The SiO(2-1)(86.847 GHz) emission is found to be an excellent indicator of active outflows from young stellar objects \citep{2007ApJ...663..1092,2014MNRAS.440..1213}. In addition, \citet{2014A&A...570A...1D} confirmed that SiO(2-1) line emission is a good indicator of outflows in massive star-forming regions.
In the cold diffuse interstellar medium (ISM), shocked dust grains can sublimate, and frozen silicon is released into the gas phase to form SiO. It can either freeze out back onto dust grains or oxidize and form SiO$_{2}$ on a time scale of 10$^{4}$ years \citep{1997IAUS..182..199P,2007ApJ...663..1092}. As emission from \hcop may potentially trace remnant, momentum-driven, 'fossil' outflows \citep{2007ApJ...663..1092}, SiO may be used as a tracer of active outflow (jet outflow).
The infall and outflow events in massive star-forming regions are closely connected to the turbulence in the ISM, which plays a dominant role in regulating massive star formation \citep{2013MNRAS.436..1245,2015MNRAS.453..3245}.
Because outflows from massive stars may contribute to driving the turbulence in the ISM (e.g., \citealt{2013MNRAS.434..2313,2014ApJ...790...128}), their feedback helps to address two main questions: (a) do outflows inject enough momentum to maintain the turbulence, and (b) can outflows couple with the clump gas and drive turbulent motions \citep{2014prpl.conf..451F}.

In this study, we identified 188 outflow candidates among 694 clumps from a previous infall survey \citep{2015MNRAS.450..1926,2016MNRAS.461..2288}.
Outflow candidates were identified from \hcop(1-0) PV diagrams. The remainder of this paper is organized as follows. A brief introduction to the survey and our sample selection is given in Section \ref{sec2}.
In Section \ref{sec3}, we identify 188 high-mass outflows and calculate their outflow parameters. Section \ref{sec4} discusses the relationship between outflow and infall and matches our sample with maser surveys to check the relationship of masers and outflows. Finally, we discuss the contribution of outflows to the turbulence in the ISM and consider the influence of outflows on the velocity dispersion at different evolutionary stages of star-formation regions. We give a summary in Section \ref{sec5}.

\section{ARCHIVAL DATA and OUR SAMPLE}\label{sec2}

\subsection{ MALT90 Survey}
The Millimetre Astronomy Legacy Team 90 GHz survey  (MALT90) aims to characterise the physical and chemical evolution of high-mass star-forming clumps \citep{2013PASA...30...57J}.
The unique broad frequency capability and fast-mapping capabilities of the Australia Telescope National Facility Mopra 22 m single-dish telescope has been exploited to simultaneously map 16 molecular lines near 90 GHz for each target source (size of each map is 3 arcmin $\times$ 3 arcmin).
MALT90 contains over 2000 dense cores identified in the APEX Telescope Large Area Survey of the Galaxy (ATLASGAL) of 870 $\mu$m continuum emission covering the Galactic plane in the longitude range of -60\degr to +20\degr \citep{2009A&A...504...415}.
These dense cores span the complete range of evolutionary stages of high-mass star formation from Pre-stellar to Proto-stellar to H\,\RNum{2} regions, and finally to photodissociation (PDR) regions  \citep{2013PASA...30...57J}.
The spatial and spectral resolution of the MALT90 survey are approximately 36 arcsec and 0.11 kms$^{-1}$. The typical rms noise of the antenna temperature is $\sigma$ = 0.25 K per channel of 0.11 kms$^{-1}$.
 We used the \hcop(1-0)(89.189 GHz) and SiO(2-1)(86.847 GHz) emission lines of MALT90 sources to identify outflows and the
N$_{2}$H$^{+}$(1-0)(93.174 GHz) emission lines from the survey to derive the full width half maximum (FWHM) line width and the system velocity of clumps.

\subsection{Methanol MultiBeam Survey}
Methanol masers are well-known indicators of the early phases of high-mass star formation \citep{1991ApJ...380...L75,2003A&A...403..1095}. \citet{2014MNRAS.444..566D} found a total of 58 $^{13}$CO(3-2)(330.588 GHz) emission peaks in the vicinity of these maser positions and found evidence of high-velocity gas in all cases.
The Methanol MultiBeam (MMB) survey mapped the Galactic plane for 6.7 GHz Class-II masers using a 7-beam receiver on the Parkes radio telescope with a sensitivity of 0.17 Jy beam$^{-1}$ and a half-power beamwidth of 3.2 arcmin \citep{2009MNRAS.392...783}.  Subsequent Australia Telescope Compact Array (ATCA) observations (in the 6-km configuration) provided high resolution positions with an accuracy of $\sim$ 0.4 arcsec \citep{2010MNRAS.404..1029}.
The MMB is complete in the range 186\degr $< \ell <$ 20\degr and $|b| <$ 2\degr (\citealt{2010MNRAS.404..1029,2011MNRAS.417..1964,2010MNRAS.409...913,2012MNRAS.420..3108}).
The MMB catalogue provides the velocity of the peak component and the flux density as measured from both the lower-sensitivity Parkes observations and the high-sensitivity ATCA follow-up observations.

\subsection{H$_{2}$O Southern Galactic Plane Survey}
Water masers usually trace outflows (e.g., \citealt{1992A&A...255...293}) at a generally earlier evolutionary phase than OH masers \citep{1989A&A...213...339}. Our aim is to test an evolutionary sequence for water and Class II methanol masers. The H$_{2}$O Southern Galactic Plane Survey (HOPS) was carried out at 22 GHz with the Mopra Radio Telescope with a broad-band backend and a beam size of approximately 2 arcmin \citep{2011MNRAS.416.1764W}.
The root mean square (rms) noise levels are typically between 1 and 2 Jy with 95 percent under 2 Jy. HOPS found 540 H$_{2}$O masers in the range 290\degr $< \ell <$ 30\degr and $|b| <$ 0.5\degr.
Subsequent high resolution observations with the ATCA show a large range of noise levels from 6.5 mJy to 1.7 Jy, but with 90 percent of noise levels in the range of 15 to 167 mJy \citep{2014MNRAS.442..2240}.
The ATCA observations had a beam size ranging from 0.55 $\times$ 0.35 arcsec to 14.0 $\times$ 10.2 arcsec.

\subsection{Sample Selection}
\begin{figure*}
\includegraphics[width=0.5\textwidth]{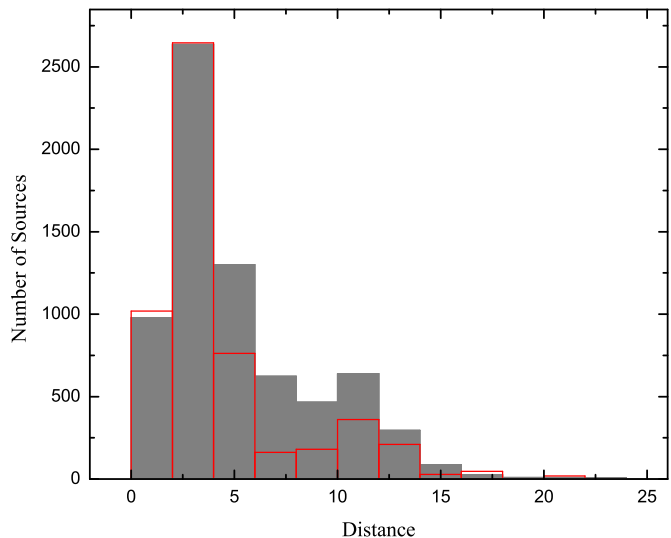}
\caption{ The distance distribution of 694 clumps (red histogram) and 8016 ATLASGAL clumps (grey filled histogram). The distance distribution of 694 clumps has been scaled to The distance distribution of 8016 ATLASGAL clumps.}\label{figa1}
\end{figure*}

\citet{2015MNRAS.450..1926,2016MNRAS.461..2288} selected 732 high-mass clumps with N$_{2}$H$^{+}$(1-0),
HNC(1-0), and \hcop(1-0) emission lines detected from the MALT90 survey with S/N $>$ 3 and an angular distance
between any two sources larger than the Mopra beam size (36 arcsec at 90 GHz). These clumps with detected N$_{2}$H$^{+}$,
HNC, and \hcop emission lines with S/N $>$ 3 are more evolved than 8016 ATLASGAL clumps at corresponding evolutionary stages. The distance range is from 0.5 kpc to 17 kpc. In Figure \ref{figa1}, our sources show slightly different from all ATLASGAL sources in distance (from 4 kpc to 10 kpc).
Thirty-eight clumps were excluded because they are close to the edge of the 3 arcmin $\times$ 3 arcmin map, which may affect the reliability of the outflow identification. Therefore, our sample included 694 massive star-forming regions. Then, we searched for outflow candidates in these 694 high-mass clumps. Since the noise is higher at the edges of the maps, a cropped 14 $\times$ 14 pixel size (approximately 2 arcmin $\times$ 2 arcmin) centred on the Galactic coordinates was selected for the ATLASGAL sources, which may results in an underestimate of the outflow parameters.
\section{RESULTS}\label{sec3}

\subsection{Outflow Identification}

\begin{figure*}
\includegraphics[width=0.47\textwidth]{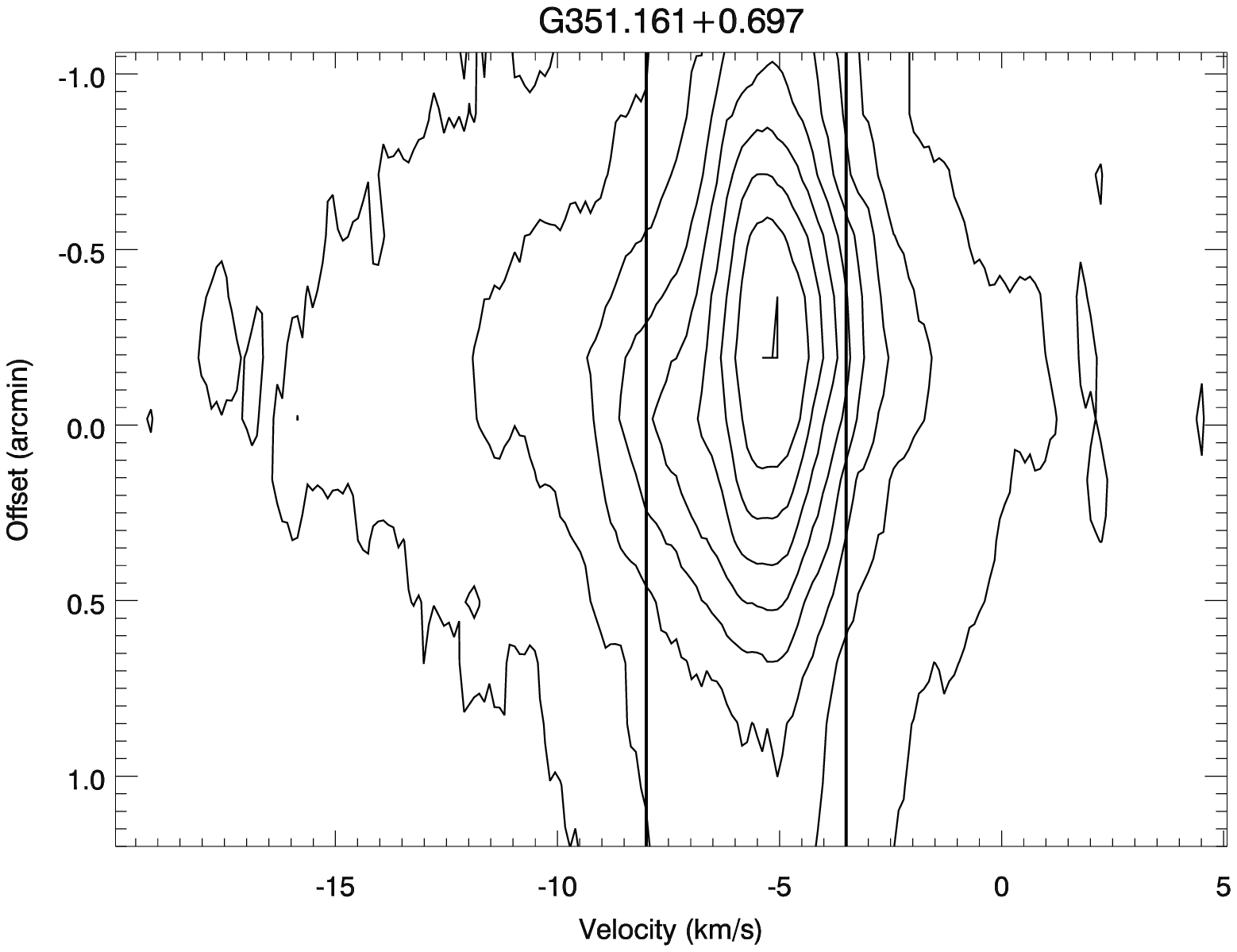}
\includegraphics[width=0.47\textwidth]{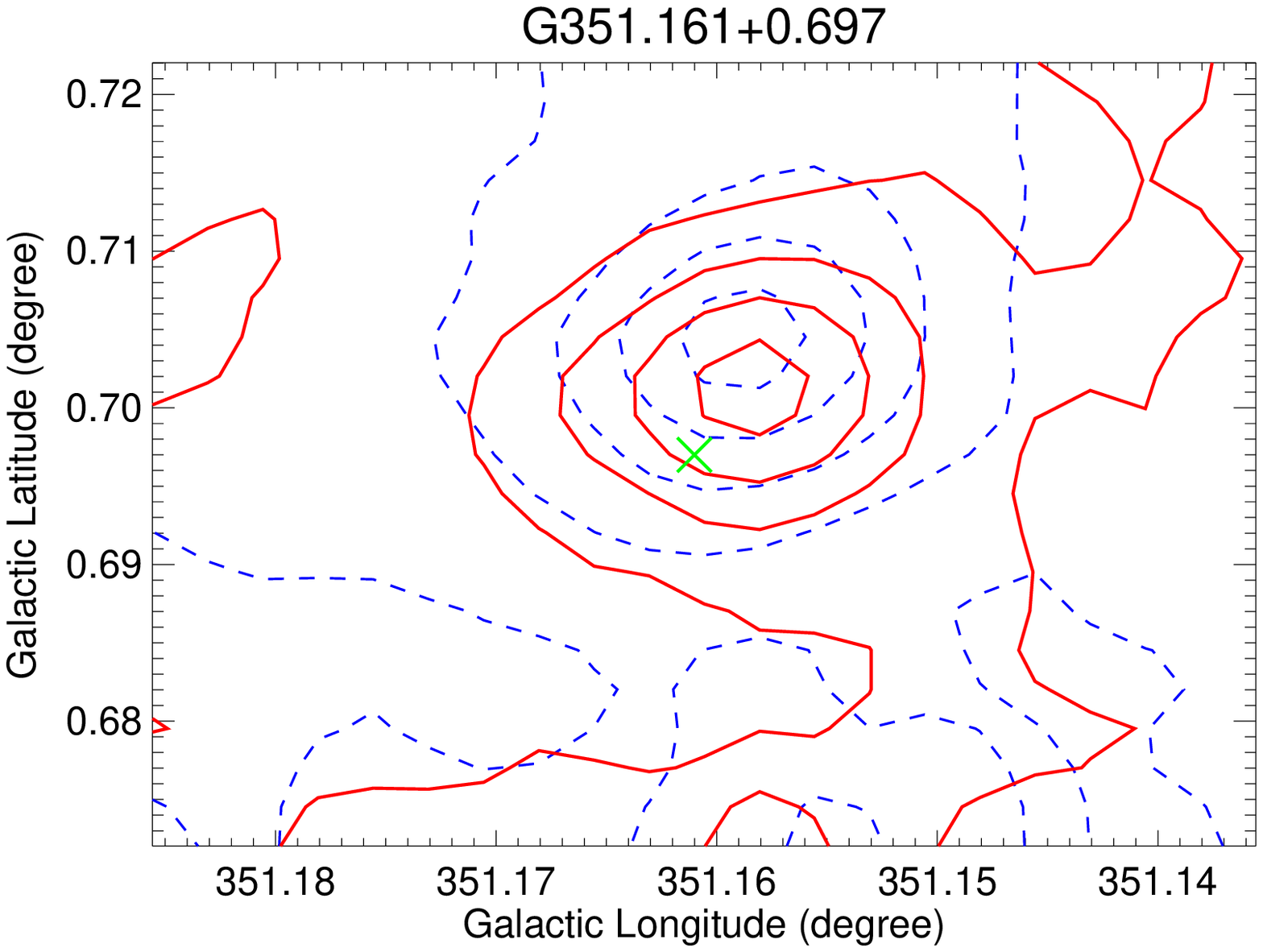}\\
\caption{{\it Left upper panel}: The PV diagram of the HCO$^{+}$ molecular emission with outflow detection for the source G351.161+0.697. Contour levels are from 3 RMS to 1.675 K in increments of 0.167 K. The left and right vertical lines indicate the beginning of the blue and red wings, respectively.
{\it Right upper panel}: The integrated intensity image of the blue (blue dashed line) and red (red solid line) wings centred on the green cross symbol at the source G351.161+0.697.
Five contour levels increase linearly from three times the RMS noise to the maximum: for blue-shifted emission, the minimum contours and contour spacings are 0.281 and 3.661 $K$\kmss, respectively, and for the red-shifted emission, they are 0.201 and 1.122 $K$\kmss.
The integrated velocity range is listed in Table \ref{tab1}.}\label{fig1}
\end{figure*}
\begin{figure*}
\includegraphics[width=0.5\textwidth]{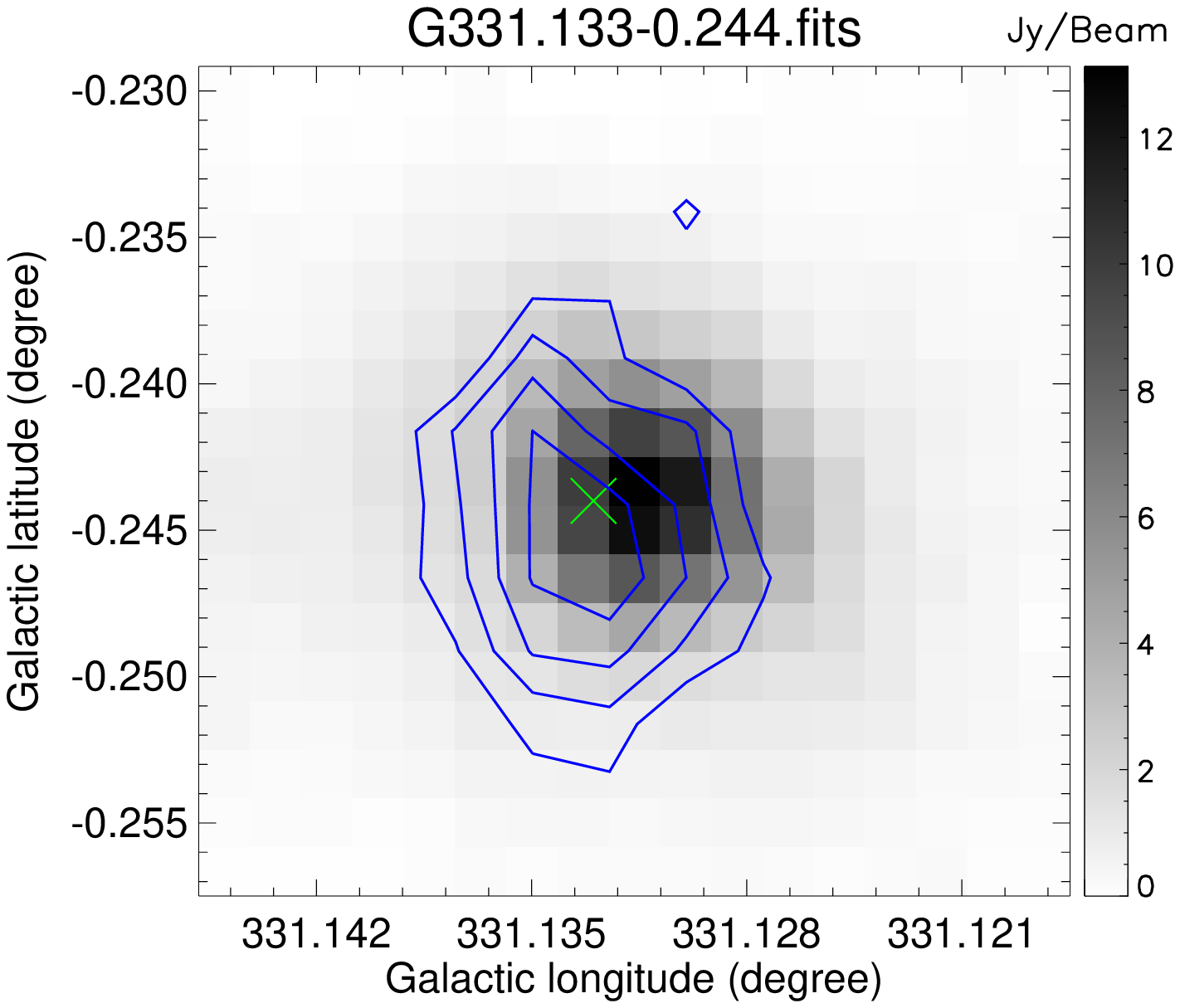}

\caption{ SiO emission integrated intensity contour superposed on 870$\mu$m continuum emission (grey scale) for the source G331.133-0.244. The centroid of the ATLASGAL clump is marked with a green cross. Four contour levels increase linearly from three times the RMS noise to the maximum. The integrated intensity map of 85 clumps with stronger SiO emission are shown online.}\label{fig2}
\end{figure*}

For each source in the sample, the identification of outflows has been made by checking extended wings representative of outflows in the PV diagrams with a cut along the Galactic latitude and longitude. We only check PV diagrams at two vertical directions because the resolution of the telescope was too low to adequately determine the outflow axis. Finally, we choose the direction with more extended wing. When the velocity bulge appears in the direction away from the central velocity at least at a 3 RMS level, we consider there is outflow, and vice versa. This step was completed by visual inspection. This may result that the reliability of some outflow candidates may be low. High-mass outflow candidates were identified in this manner. An example of a high-mass outflow candidate in an \hcop PV diagram is displayed in Figure \ref{fig1}.The contour levels of all clumps start from 3 sigma. The PV diagrams of 188 clumps are showed online.
 From the 188 outflow candidates, 43 sources had only one clear lobe, the red or blue lobes of 24 sources were contaminated by the gas along the line of sight, 13 sources had overlapping blue and red lobes, and 3 sources had no distance estimate. Finally, this leaves a sample of 105 clumps with suitable bipolar outflows and reliable distances for which the outflow parameters may be calculated. In addition, the SiO(2-1) emission in our sample clumps was used as an indicator of active outflows in massive star-forming regions, where the detected SiO emission is caused by outflow-driven shocks \citep{2014A&A...570A...1D,2014A&A...562....A3}.
A total of 198 sources were found with SiO emission above 3 sigma in integrated intensity and 85 clumps of these with stronger SiO emission also show HCO$^{+}$ line wings. The SiO emission integrated intensity maps of 85 clumps with stronger SiO emission are presented online and an example of an SiO detection is shown in Figure \ref{fig2}.

\subsection{Outflow Parameters}
We determine the velocity ranges of high-velocity gas emission from PV diagrams and choose 50\% of peak emission as the beginning of the high-velocity wings, at which a velocity gradient clearly begins and is spatially extended. Assuming that the HCO$^{+}$ emission in the line wings is optically thin and in local thermodynamic equilibrium (LTE), the following values may be adopted for the abundance ratio and the excitation temperature of $[H_{\rm 2}/HCO^{+}] = 10^{8}$  and $T_{ex}$ = 15K \citep{1997ApJ...483..235T}.
The physical properties of the outflows may then be calculated following the procedure of \citet{1991ApJ...374..540G,2015ApJS..219...20L}.
The total column density of the outflow gas is given by
\begin{equation}
N(HCO^{+}) = \frac {3k^2T_{\rm ex}}{4\pi^3\mu^2_{\rm d}h\nu^2exp(-h\nu/kT_{\rm ex})}\int T_{\rm mb}d\upsilon
\end{equation}
where the Boltzmann constant $k$ = 1.38 $\times$ 10$^{-16}$ erg K$^{-1}$, the Planck constant $h$ = 6.626 $\times$ 10$^{-27}$ ergs, the dipole moment $\mu_{\rm d}$ = 3.89 $\times$ 10$^{-18}$ esu cm$^{-1}$, the transition frequency $\nu$ = 89.188526 GHz, and the velocity $v$ is in km s$^{-1}$.
T$_{\rm{mb}}$ is the brightness temperature, which is calculated from the antenna temperature, dividing it by a main beam efficiency of 0.49 \citep{2013PASA...30...57J}.
The mass for each pixel in the defined outflow lobe area is computed by
\begin{equation}
M_{\rm pixel} = N(HCO^{+})[H_{\rm 2}/HCO^{+}]\mu_{\rm H_{\rm 2}}m_{\rm H}A_{\rm pixel},
\end{equation}
where the mean molecular weight $\mu_{\rm H_{\rm 2}}$ = 2.72 \citep{2010A&A...513A..67B}, the mass of the hydrogen atom $m_{\rm H}=1.67 \times$ 10$^{-24}$ g, and the area of each pixel $A_{\rm pixel}$ within the outflow lobe is contained within the 3 sigma contours integrated intensity in \hcop.

The total mass, the momentum (P$_{out}$) and the kinetic energy (E$_{out}$) of each outflow is obtained by summing over all spatial pixels defined by the lowest contours. Finally, the mass rate of the outflow, the mechanical luminosity, and the mechanical force are calculated as $\dot{M}_{\rm out} = M_{\rm out}/t_{dyn}$,$L_{\rm out} = E_{\rm out}/t_{dyn}$, and $F_{\rm out} = P_{\rm out}/t_{dyn}$, respectively. The details refer to \citet{li}.
 The dynamical timescale can be calculated by two methods. One is to calculate the ratio of the maximum separation of the outflow lobes (the length of the blue and red lobes) and the terminal speed of the outflow measured from the \hcop spectral line wings. The other method is to calculate the ratio of the separation between the peak of the blue and red lobes (the distance between the lobes) and the mean outflow velocity defined as $P_{\rm out}/M_{\rm out}$. We adopt the second method because the first method requires higher observational sensitivity. \citet{1999ApJ...522..921G} indicates that in the optically thin limit, $T_{\rm mb}(4-3)/T_{\rm mb}(1-0)$ will yield a good approximation of the excitation temperature when it is less than 15 K and a lower limit when it is higher than this value. Our choice of $T_{\rm ex}(J=1-0)$=15 K implies that the \hcop emission in the line wings is optically thin. This excitation temperature is a lower limit when the opacity increases. The outflow mass (and all other properties that depend on mass) reaches a smallest value at an excitation temperature 5 K. When the excitation temperature is more than 5 K, the outflow properties only increase with increasing excitation temperature and the outflow mass (and all other properties that depend on mass) will be a lower limit when the opacity increases. Many of the uncertainties in distance, \hcop abundance, and inclination angle \citep{1990ApJ...348...530}  are systematic and have little effect on the overall distribution and correlations between the individual quantities. Consequently, the homogeneity of our sample and the large number of objects will ensure robust results from our statistical analysis. The sensitivity is unlikely to significantly affect the total $M_{\rm out}$ because of the steeply declining nature of the outflow mass spectra \citep{2014ApJ...783....29}. The outflow mass would be overestimated when the low-velocity outflow emission contains some ambient cloud emission.
We adopt an average inclination angle of $\theta$ = 57.3\degr to correct the results \citep{2015ApJS..219...20L,2015MNRAS.453..3245}. The correction factors of the momentum and the kinetic energy of the outflows are 1.9 and 3.4, respectively.
The inclination-corrected physical properties of the outflows are partly listed in Table~\ref{tab1}, and the whole list is available online as Table A1. The ratio of $M_{\rm out}/M_{\rm clump}$ has an average of 0.03 for the sources and a spread of less than one order of magnitude, which is similar to the mean ratio of 4\% in \citet{2002A&A...383..892B} and 5\% in \citet{2018ApJS..235.....3}.

\begin{table*}
\scriptsize
 \centering
  \begin{minipage}{185mm}
\caption{Examples of the outflow properties of the blue and red lobes: masses M$_{out}$, momentum p, energy E, dynamic time t$_{dyn}$, mechanical force F$_{out}$, mechanical luminosity L$_{out}$, mass-loss rates $\dot{M}_{out}$, velocity range $\Delta$$V_{b}$, $\Delta$$V_{r}$, distance, and distance uncertainty. \label{tab1}}
\begin{tabular}{lcccccccccc}
\hline
\multicolumn{1}{c}{Clump$^{a}$} &
\multicolumn{1}{c}{M$_{out}$} &
\multicolumn{1}{c}{p} &
\multicolumn{1}{c}{E}&
\multicolumn{1}{c}{t$_{dyn}$}&
\multicolumn{1}{c}{F$_{out}$}&
\multicolumn{1}{c}{L$_{out}$} &
\multicolumn{1}{c}{$\dot{M}_{out}$} &
\multicolumn{1}{c}{$\Delta$$V_{b}$} &
\multicolumn{1}{c}{$\Delta$$V_{r}$}&
\multicolumn{1}{c}{Distance(error)}\\
\multicolumn{1}{c}{name}&
\multicolumn{1}{c}{($M_{\bigodot}$)}&
\multicolumn{1}{c}{($M_{\bigodot}$\kms)}&
\multicolumn{1}{c}{($10^{45}erg$)}&
\multicolumn{1}{c}{($10^{4}yr$)}&
\multicolumn{1}{c}{($10^{-3}M_{\bigodot}kms^{-1}yr^{-1}$)}&
\multicolumn{1}{c}{($L_{\bigodot}$)}&
\multicolumn{1}{c}{($10^{-4}M_{\bigodot}yr^{-1}$)}&
\multicolumn{1}{c}{(\kms)}&
\multicolumn{1}{c}{(\kms)}&
\multicolumn{1}{c}{kpc}\\
\hline
G000.316-0.201 & 451.1 & 2272.3 &  239.53 &19.3 & 17.97 & 9.611  & 23.37   & [13.8,16.8]  &  [20.3,23.4] & 8.0(0.12) \\
G000.546-0.852 &  58.3 &  414.5 &   61.26 & 8.5 &  7.44 & 5.582  &  6.86   & [10.5,13.8]  &  [18.5,24.5] & 2.0(0.12) \\
G005.899-0.429 &  22.4 &   87.9 &    6.93 &45.8 &  0.29 & 0.117  &  0.49   &  [2.9,4.4]  &   [8.4,9.8] & 2.7(0.15) \\
G005.909-0.544 &  23.8 &  101.7 &    8.69 & 3.7 &  4.19 & 1.819  &  6.44   & [12.0,13.9]  &  [17.0,19.0] & 3.27(0.15) \\
G006.216-0.609 &  18.4 &   78.0 &    6.38 &11.2 &  1.06 & 0.441  &  1.64   & [14.8,16.1]  &  [19.5,20.8] & 3.58(0.15) \\
\hline
\end{tabular}
\\$^{\rm{a}}$ Sources are named by galactic coordinates of the maximum intensity in the ATLASGAL sources.
\end{minipage}
\end{table*}

\section{DISCUSSION}\label{sec4}

\subsection{Detection Statistics of Outflows}
From the 694 sources a total of 188 outflow candidates, resulting in a detection rate of 27\%. The detection rate is smaller than  found in previous studies (\citealt{2018ApJS..235.....3} (66\%), \citealt{2015MNRAS.453...645} (66\%), \citealt{2001ApJ...552L.167Z,2005ApJ...625..864Z} (57\%), and \citealt{2004A&A...417...615} (39\%$\sim$50\%)) but is similar to the 20\% detection rate found in \citet{li}.
A lower detection rate is expected for sources located in the inner region of the Galactic plane with higher interstellar extinction and internal absorption.

Among the 694 clumps in our sample, there are 61 pre-stellar sources, 278 proto-stellar sources, 230 H\,\RNum{2} regions, 66 photo dissociation regions (PDRs), and 59 with an uncertain classification.
These classifications were obtained from \citet{2015ApJ...815..130G} and \citet{2016MNRAS.461..2288}. The classification is based on  3.6, 4.5, 8.0, and 24 $\mu$m \textit{Spizter} images (see \citet{2015ApJ...815..130G} for details). We note that there are 3 outflow candidates at a pre-stellar stage but these may be misclassified. We classify these as proto-stellar. So there are 58 pre-stellar sources and 281 proto-stellar sources.
Outflow line wings were detected using HCO$^{+}$ towards 75 proto-stellar sources (75/281 or 27\%), 88  H\,\RNum{2} regions (88/230 or 38\%), and 17 PDRs  (17/66 or 26\%).
This indicates that outflow detection rate increases from the proto-stellar to H\,\RNum{2} evolutionary stage. If the outflow is just switched off, the detection rate of \hcop outflow should be constant with a similar detection rate as that of the previous stage. However, the lower detection rate at the PDR stage suggests that the previous jet entrained gas was not detected and may have likely been blown away. The decreased detection rate at the PDR stage is likely because the circumstellar envelope matter was blown away.

Similarly among the 188 outflow candidates, the SiO emission was detected towards 36 proto-stellar sources (36/75 or 48\%), 42 H\,\RNum{2} regions (42/88 or 48\%), and 3 PDR regions (3/17 or 18\%). We select SiO emission as a tracer of active outflow (not a 'fossil' outflow). If there is a stellar wind, both the \hcop and SiO gas would be blown away to a lower column density and would not be etected.
The sharply decreased detection rate of \hcop and SiO from H\,\RNum{2} to PDR supports that outflows have mostly been switched off, and that the circumstellar envelope matter has been blown away during the PDR stage.

\subsection{Outflows and Masers}

A search of masers within a beam for each clump shows that among the 694 clumps there are 123 clumps associated with maser sources: 82 clumps with methanol masers and 62 clumps with water masers.
Sixty-two of these 123 clumps are outflow candidates which makes a high outflow detection rate (50\%). This suggests an intimate relationship between outflow action and the presence of masers. Forty methanol masers (40/82; 49\%) and thirty-four water masers are outflow candidates (34/62; 55\%).
Among the 82 clumps associated with the methanol masers, there are 1 pre-stellar source, 37 proto-stellar sources, 40 H\,\RNum{2} regions, 3 PDR sources, and 1 source with uncertain classification. The outflows are detected towards 13 proto-stellar (13/37; 35\%), 25 H\,\RNum{2} regions (25/40; 63\%), and 1 PDR source (1/3; 33\%).
Among the 62 clumps associated with water masers, there are 1 pre-stellar source, 23 proto-stellar sources, 34 H\,\RNum{2} regions, 1 PDR source, and 3 sources with uncertain classification. The outflows are detected towards 10 proto-stellar (10/23; 43\%), 22 H\,\RNum{2} regions (22/34; 65\%), and 0 PDR source (0/1; 0\%).
 Similar outflow detection rate towards methanol masers and water masers at proto-stellar and H\,\RNum{2} stage possibly mean that water masers appear at a nearly similar stage as 6.7 GHz methanol masers. We note that there are less samples at the PDR stage. Therefore, the detection rate may be not accurate at the PDR stage.

\begin{figure}
\includegraphics[width=0.5\textwidth]{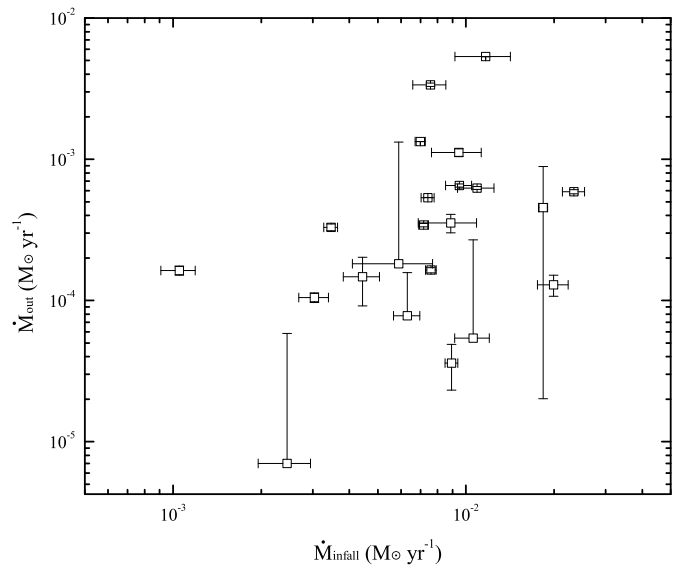}
\caption{ The outflow mass rate ($\dot{M}_{\rm out}$) of clumps compared with their infall mass rate from the literature. The uncertainties are based on the distance uncertainty.}\label{fig3}
\end{figure}

\subsection{Outflow and Infall}
\citet{2015MNRAS.450..1926,2016MNRAS.461..2288} use the optically thin line of N$_{2}$H$^{+}$ and the optically thick lines of HNC and HCO$^{+}$ for searching for infall candidates. By calculating an asymmetry parameter, they determined red and blue skewed profiles of the optically thick lines of each clump. Infall candidates must show a blue skewed profile at least in one optically thick line, no red skewed profile in the other optically thick line and no spatial difference in the mapping results.

Among the 694 clumps, there were 222 infall candidates (infall candidates refer to \citet{2015MNRAS.450..1926,2016MNRAS.461..2288}). Infall is detected in 35\% (178 out of 506) of 506 clumps without outflow. However, infall is detected in only 23\% (44 out of 188) of 188 outflow candidates. This suggests that the infall detection rate is lower towards outflow candidates.
 Among the outflow candidates, 31\% of proto-stellar clumps have evidence of infall (23 out of 75), 23\% of H\,\RNum{2} regions have evidence of infall (20 out of 88), and 0\% of PDR clumps have evidence of infall (0 out of 17).
 Among non-outflow candidates, 40\% of proto-stellar clumps have evidence of infall(82 out of 206), 37\% of H\,\RNum{2} regions have evidence of infall (52 out of 142), and 16\% of PDRs have evidence of infall (8 out of 49).
 The infall detection rates towards outflow candidates is always lower at the corresponding stage. This indicates that outflow action decreases the infall detection rate at each evolutionary stage. Outflows entraining the surrounding gas may have some effect on the infall process. The infall detection rates towards non-outflow candidates are rather constant from the proto-stellar to the H\,\RNum{2} stages. However, the infall detection rates towards outflow candidates show a larger change from the proto-stellar to H\,\RNum{2} regions. The infall detection rate towards non-outflow candidates is constant from proto-stellar to H\,\RNum{2}. If the outflow effect is constant, then the infall detection rate towards the outflow candidates should be constant, but it is not. This suggests that the effect of outflows on their environment is becoming more significant.

 We find that infall is detected in 28\% of 85 sources with SiO emission (evidence of jets) and \hcop outflow detections.
 Among the 85 sources with SiO emission and outflow detections, evidence of infall is found in 33\% of the proto-stellar clumps (12 out of 36), 26\% of the H\,\RNum{2} regions (11 out of 42) and in none of the 3 PDRs. It seems that there is no significant difference in the infall detection rate towards outflow candidates with or without SiO emission at corresponding evolutionary stages.
 Since candidates are identified by a variable line of sight velocity, outflow and infall is detected in a parallel direction.
 Therefore, outflow action produces some effects on the infall process for both jet or 'fossil' outflows.

We calculated the outflow parameters of 105 clumps including the 22 infall candidates and the infall mass rate was obtained from \citet{2015MNRAS.450..1926,2016MNRAS.461..2288}. Figure \ref{fig3} shows the outflow mass rate ($\dot{M}_{\rm out}$) versus the  infall mass rate for the 22 sources. When comparing the outflow mass rate of the clumps with the infall mass rate, there is no clear relationship between the two (Spearman's rank correlation: $\rho$ = 0.3, p-value = 0.18).
The basic information of part of the clumps is listed Table \ref{tab2}, and the whole list is available online as Table A2.
\begin{table*}
\caption{The basic information of part clumps: clump mass Log(M$_{clump}$), local region FWHM of N$_{2}$H$^{+}$, clump FWHM of N$_{2}$H$^{+}$, evolution type, and notes. Here, "o" = outflow candidates, "i" = infall candidates, "w" = water masers, "m" = methnal masers, and "s" = SiO. Sources are named by galactic coordinates of the maximum intensity in the ATLASGAL sources. Infall candidates refer to \citet{2016MNRAS.461..2288}. \label{tab2}}
\begin{tabular}{lccccc}
\hline
\multicolumn{1}{c}{Clump} &
\multicolumn{1}{c}{Log(M$_{clump}$)}&
\multicolumn{1}{c}{Pixel FWHM}&
\multicolumn{1}{c}{FWHM}&
\multicolumn{1}{c}{Stage}&
\multicolumn{1}{c}{Notes} \\
\multicolumn{1}{c}{name}&
\multicolumn{1}{c}{M$_{\bigodot}$}&
\multicolumn{1}{c}{\kms}&
\multicolumn{1}{c}{\kms}&
\multicolumn{1}{c}{}&
\multicolumn{1}{c}{}\\
\hline
G010.288-00.124 &   2.76       &   2.74    &    4.136    &   HII          &  i   s \\
G010.299-00.147 &   3.37       &   3.38    &    4.579    &   HII          &   o  s \\
G010.323-00.161 &   3.44       &   2.29    &    3.027    &   HII          &    wm  \\
G010.329-00.172 &   4.07       &   1.97    &    2.799    &   PDR          &        \\
G010.342-00.142 &   2.94       &   2.7     &    3.837    &   Proto-stellar&   owms \\
\hline
\end{tabular}
\end{table*}

\subsection{Clump turbulence}\label{lab41}

\begin{figure}
\includegraphics[width=0.5\textwidth]{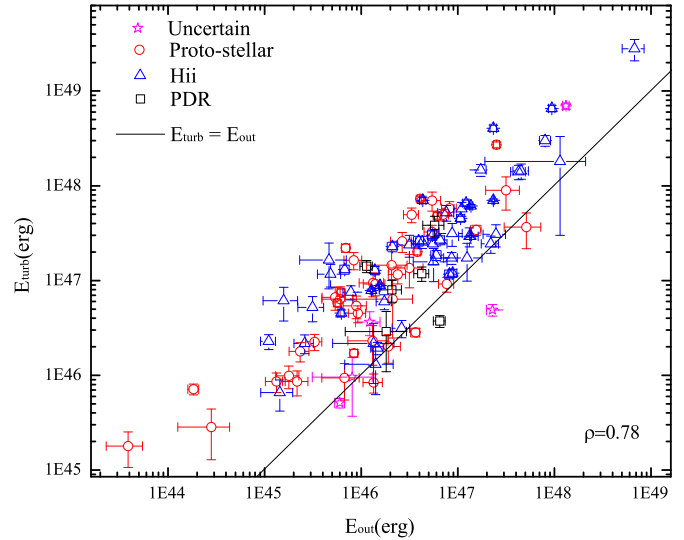}
\caption{ The turbulent energy in clumps compared with the outflow energy.
The magenta stars, red circles, blue triangles, and black squares refer to sources with uncertain, proto-stellar, H\,\RNum{2}, and PDR classification, respectively.
The solid line shows E$_{out}$ = E$_{turb}$. The uncertainties are based on the distance uncertainty.}
\label{fig4}
\end{figure}
The presence of outflows in clumps may have a cumulative impact on the level of turbulence in molecular clouds.
One method to quantify this effect is to compare the total energy of the outflow with the cloud's total turbulent kinetic energy. This total turbulent energy may be estimated as E$_{turb}$=(3/16 $\ln2$)M$_{cloud}$$\times$FWHM$^{2}$ \citep{2001ApJ...554...132}. Assuming that the N$_{2}$H$^{+}$ gas temperature is equal to the dust temperature in LTE, we estimate the expected thermal motions of N$_{2}$H$^{+}$ to show that it is insignificant in the FWHM. The average of the ratios between the velocity dispersion (thermal motions) and total velocity dispersion is 0.12, which means that the thermal motions have little contribution to the FWHM.
A comparison of the turbulent energy and the outflow energy of our sample sources in Figure \ref{fig4} shows that the outflow energy correlates with and is comparable to the turbulent energy (Spearman's rank correlation coefficient 0.83 and p-value $\ll$ 0.001).
This indicates a strong relationship between the turbulent energy and the outflow energy in the source.

\begin{figure}
\includegraphics[width=0.5\textwidth]{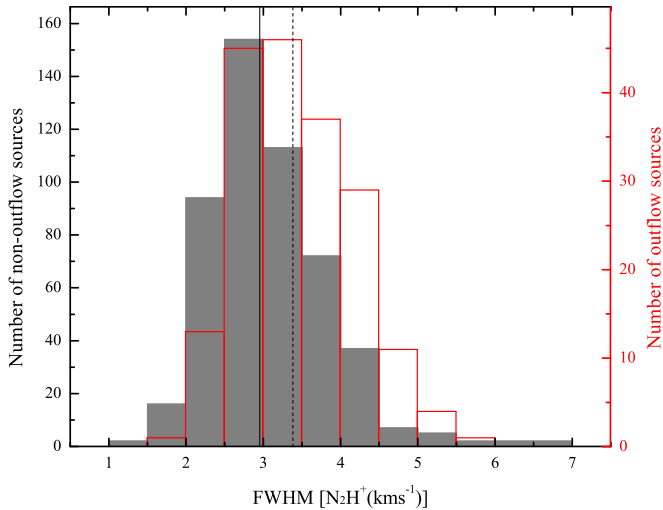}
\caption{ The N$_{2}$H$^{+}$ FWHM distribution of clumps with detected outflows (red histogram) and without detected outflows (grey filled histogram). The dashed vertical black line and the solid vertical black line are the median values of N$_{2}$H$^{+}$ FWHM for clumps without and with outflow, respectively. The FWHM distribution of clumps with detected outflows has been scaled to FWHM distribution of clumps without detected outflows.}\label{fig5}
\end{figure}

Because there is no correlation between the outflow and the clump mass, the effect of distance may be excluded and we just need to consider whether the outflow has a significant effect on the clump FWHM and hence on the turbulent energy.
As a first test, the FWHM distribution of the N$_{2}$H$^{+}$ lines are compared for the clumps with and without outflows in Figure \ref{fig5}. This shows that the outflow candidates show a slightly higher median value for the FWHM than the clumps without detected outflows, which suggests that outflows do have a relatively significant effect on the N$_{2}$H$^{+}$ FWHM of the clump and hence on the turbulent energy.
A Kolmogorov-Smirnoff (K-S) test suggests that the two samples are indeed drawn from different parent distributions (statistic = 0.24 and p-value $\ll$ 0.001).

\begin{figure}
\includegraphics[width=0.5\textwidth]{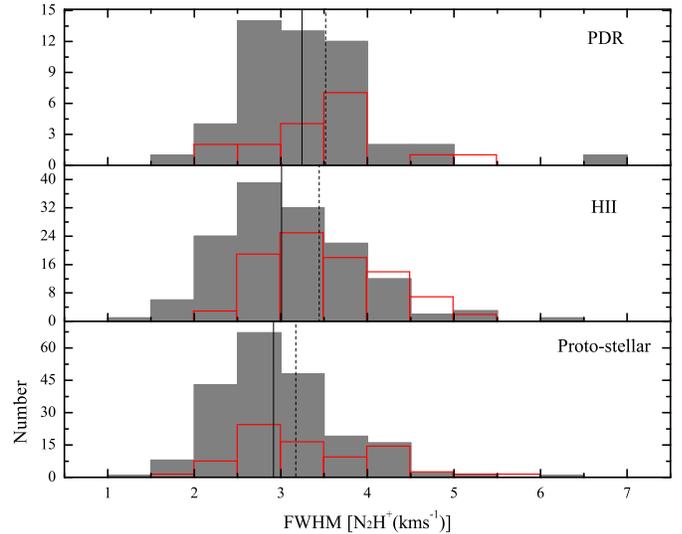}
\caption{ The N$_{2}$H$^{+}$ FWHM distributions of clumps with a detected outflow (red histogram) and without detected outflow (grey filled histogram) at each evolutionary stage. The medians for clumps with and without outflow in each stage are indicated by the dashed vertical black line and the solid vertical black line, respectively.}\label{fig6}
\end{figure}

Next, a comparison can be made between clumps with and without outflows at each corresponding stage in order to determine whether outflows have a significant effect on the clump FWHM at different evolutionary stages (Figure \ref{fig6}).
Clumps with outflows are found to have a larger median value of the N$_{2}$H$^{+}$ FWHM than clumps without outflows at each evolutionary stage. This suggests that outflows have an effect on the clump FWHM at each evolutionary stage. K-S tests for proto-stellar and H\,\RNum{2} regions, respectively, suggest that clumps with and without outflow are drawn from different parent distributions (statistic = 0.21 and p-value = 0.01 for proto-stellar clumps, statistic = 0.28 and p-value $\ll$ 0.001 for H\,\RNum{2} regions).
In order to estimate the contribution of the outflow to the FWHM of each clump, we calculate the median ratio of N$_{2}$H$^{+}$ FWHM between the clumps without and with the outflow for proto-stellar sources, H\,\RNum{2} regions, and PDRs as 0.92, 0.87, and 0.92, respectively. The median ratio represents the non-outflow contribution
(non-outflow contribution/all contributions; the N$_{2}$H$^{+}$ median FWHM for non-outflow detections/N$_{2}$H$^{+}$ median FWHM for outflow detections). This ratio slightly decreases between the proto-stellar and H\,\RNum{2} stages, which suggests that during this time interval the outflow contribution to the FWHM increases.
However, the ratio increases slightly from the H\,\RNum{2} to the PDR stage, which suggests that the outflow contribution to the FWHM decreases. A K-S test shows a much smaller difference and a 33\% probability that clumps with and without outflow are drawn from the same distribution for PDRs (statistic = 0.26 and p-value = 0.33), which supports that the outflow contribution may be disappearing. This may be because the outflow generating mechanism in the PDR clumps of our sample already have shut down and the observed outflow gas represents a 'fossil' outflow from previous accretion episodes. This comparison provides weak evidence that the influence of outflows on the FWHM of the clumps increases from the proto-stellar to H\,\RNum{2} stage and decreases with increasing evolutionary time when the outflow action ceases. Note that the number of pre-stellar clumps in our sample is small.
\begin{figure}
\includegraphics[width=0.5\textwidth]{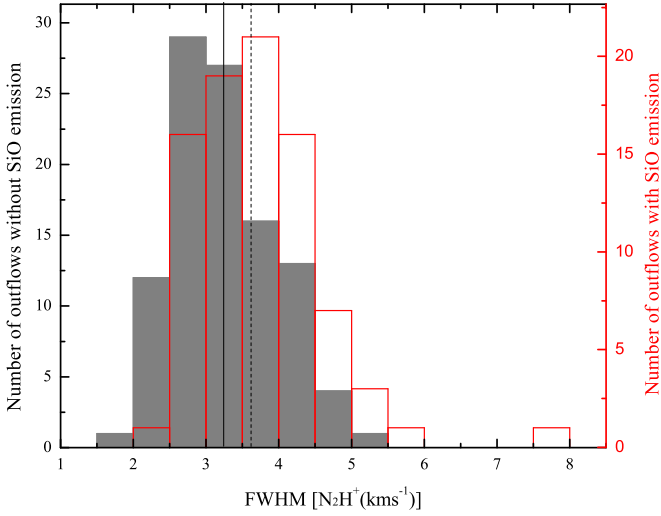}
\caption{The N$_{2}$H$^{+}$ FWHM distribution of outflow candidates with and without SiO emission.
The red histogram represents clumps with SiO emission and the grey filled histogram represents clumps without SiO emission.
The medians for outflow candidates with and without SiO emission are indicated by the dashed vertical black line and the solid vertical black line, respectively.
 The FWHM distribution of outflow with SiO emission has been scaled to FWHM distribution of outflow without SiO emission.}\label{fig7}
\end{figure}

A similar comparison of outflow candidates with and without SiO emission shows that outflow candidates with SiO emission have a larger FWHM median value, which suggests that the outflows with SiO emission contribute to the FWHM (Figure \ref{fig7}). A K-S test suggests that the two samples are drawn from different parent distributions (statistic = 0.30 and p-value $\ll$ 0.001).
SiO emission appears to be a good tracer of active outflows (no 'fossil' outflow) since outflow-driven shocks can sublimate dust grains and release the frozen silicon into the gas phase forming SiO. SiO can either freeze out back onto the dust grains or oxidize and form SiO$_{2}$ after a few 10$^{4}$ yr \citep{1997IAUS..182..199P,2007ApJ...663..1092}. Outflow candidates with SiO emission have a larger FWHM median value than outflow candidates without SiO emission, which also indicates that the outflow contribution decreases with time as the outflow action ceases. This means that outflows do not have a significantly cumulative impact on the turbulence levels.

\begin{figure}
\includegraphics[width=0.5\textwidth]{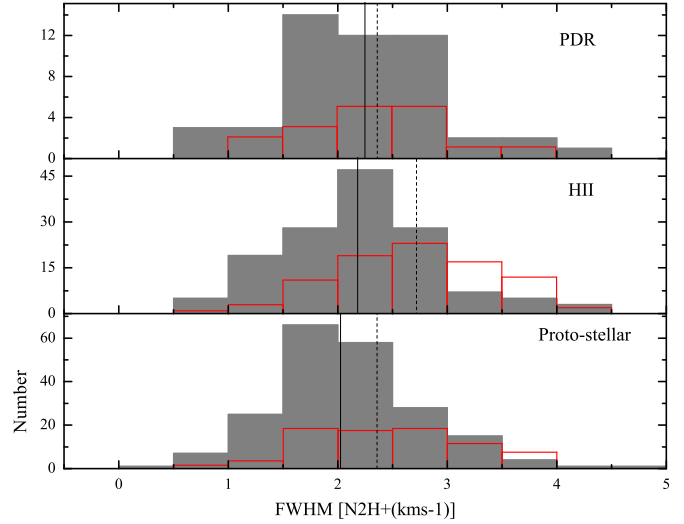}
\caption{The N$_{2}$H$^{+}$ FWHM distributions of local regions of clumps with and without outflows. The red histogram represents clumps with an outflow and the grey filled histogram represents clumps without an outflow at each stage. The medians for clumps with and without outflows at each stage are indicated by the dashed vertical black line and solid vertical black line, respectively.}\label{fig8}
\end{figure}
\begin{figure}
\includegraphics[width=0.5\textwidth]{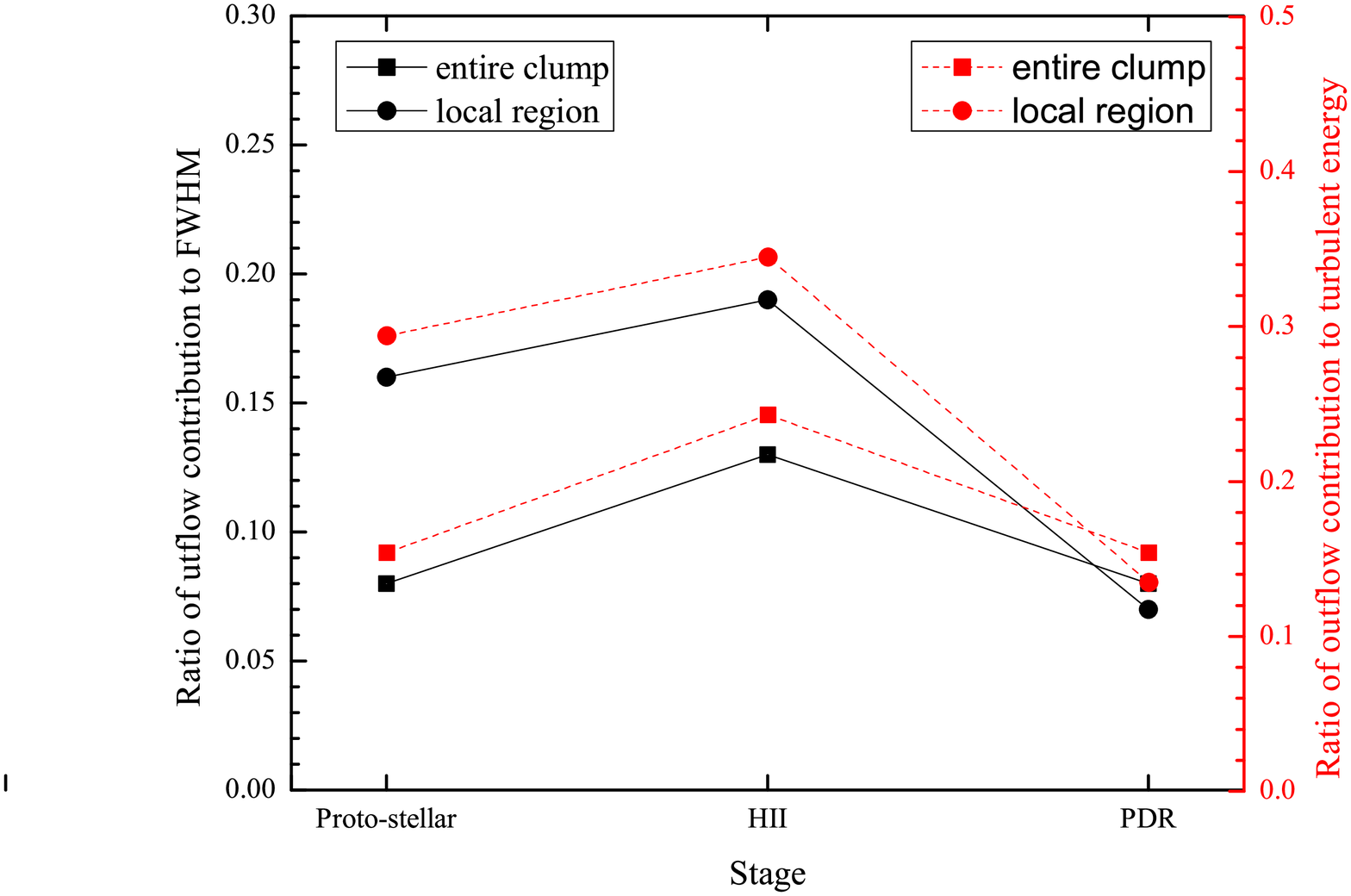}
\caption{The variation of the ratio of the outflow contribution to the FWHM (black solid line) and the turbulent energy (red dashed line). The squares denote result for entire clumps and circles indicate result for local regions.}\label{fig9}
\end{figure}

The influence of outflows on the FWHM of the local region may be determined from the FWHM of a pixel at the peak flux position in the 870 $\mu$m continuum emission (hereafter referred to as pixel FWHM).
Figure \ref{fig8} shows the distributions of the pixel FWHM of clumps with and without outflows at each evolutionary stage.
Clumps with an outflow have larger pixel FWHM median values than clumps without an outflow at corresponding evolutionary stages, and the pixel FWHM increases from the proto-stellar to the H\,\RNum{2} stage regardless of the presence of an outflow. This suggests that outflows contribute to the pixel FWHM.
Likewise, the ratios of the pixel FWHM values between clumps without and with outflow for proto-stellar sources, H\,\RNum{2} regions, and PDRs are 0.84, 0.81, and 0.93, respectively. These ratio values are relatively similar from the proto-stellar to the H\,\RNum{2} stage, which suggests that outflow contribution may be constant at these stages. K-S tests for proto-stellar, H\,\RNum{2} regions, and PDRs at local regions, respectively, show the same statistical significance as the entire clumps (statistic = 0.25 and p-value = 0.002 for proto-stellar clumps, statistic = 0.33 and p-value $\ll$ 0.001 for H\,\RNum{2} regions, statistic = 0.22 and p-value = 0.53 for PDRs). All median values are listed in Table \ref{tab3}.

 In Figure \ref{fig9}, we plot the variation of the ratio of the outflow contribution to the FWHM and turbulent energy. The ratio of the outflow contribution = 1 -- ``non-outflow contribution"/``all contributions". We observe that the outflow has a contribution in the FWHM: about 20\% in the local region at the H\,\RNum{2} region (non-outflow contribution is about 81\%) and about 10\% even in the clumps. According to E$_{turb}$=(3/16 $\ln2$)M$_{cloud}$$\times$FWHM$^{2}$, outflow has a contribution in the turbulent energy up to 35\% in the local region at the H\,\RNum{2} region (1-0.81$^{2}$). It has a contribution of at least 15\% in the clump at early stages of massive star formation, which is lower than that reported in previous studies (e.g. \citealt{2016ARA&A..54..491B,2016MNRAS.457L..84D}). The outflow contribution decreases with time once the outflow action stops. This indicates that the outflows do not have a significant cumulative impact on the turbulent levels during the occurrence of several outflow actions. Thus, the outflow energy contribution to turbulent energy increases insignificantly with the evolutionary stages. Our results suggest that the outflow energy is large enough to maintain the turbulent energy in the clumps and that the outflow has some (not significant) effect on the turbulent energy. However, there is a better correlation between the outflow energy and turbulent energy (see Figure \ref{fig4}). Therefore, we could not determine if the outflow significantly contributes to the turbulent energy in the clumps. This is consistent with the study conducted by \citet{2015MNRAS.453...645}. They also reported that there is a better correlation between the outflow energy and turbulent energy, but the core turbulence is not driven by the local input from the outflows. However, \citet{2016MNRAS.457L..84D} and \citet{2018ApJS..235.....3} reported that there is not correlation between the turbulent and outflow energies. \citet{2018MNRAS.473..1059} found that the clump mass and evolutionary stage are uncorrelated. For similar mass of massive star, it is likely that we can observe the obvious difference of turbulent energy between clump without and with outflow. However, for statistics, the mass parameter of turbulent energy is less constrained for each evolutionary stage.
All these findings imply that the outflow action has some impact on the local environment and cloud itself, but the contribution from outflow does not mainly drive turbulence. This observation is consistent with several other studies that suggest that turbulence is mostly driven by large-scale mechanisms \citep{2002A&A...390...307,2009A&A...504...883,2009ApJ...707...L153,2010ApJ...715..1170,2012MNRAS.420....10,2015ApJ...803....22,2016MNRAS.457L..84D}.

\begin{table*}
\caption{Summary of clumps with and without outflow properties. \label{tab3}}
\begin{tabular}{lccccccc}
\hline
\multicolumn{1}{c}{Property}&
\multicolumn{1}{c}{Group}&
\multicolumn{1}{c}{Notes}&
\multicolumn{1}{c}{Counts}&
\multicolumn{1}{c}{Median}\\
\hline
FWHM(N$_{2}$H$^{+}$)    &all clumps    &1&506 &2.96\\
(kms$^{-1}$)       &Proto-stellar &1&206 &2.92\\
                  &              &2&75  &3.18\\
                  &H\,\RNum{2}   &1&142 &3.01\\
                  &              &2&88  &3.45\\
                  &PDR           &1&49  &3.24\\
                  &              &2&17  &3.53\\
FWHM(N$_{2}$H$^{+}$)  & outflow candidate & without SiO emission& 103 & 3.24 \\
(kms$^{-1}$)   &                   & with SiO emission   & 85 & 3.63 \\
pixel FWHM(N$_{2}$H$^{+}$)  &Proto-stellar &1&206 &2.03\\
(kms$^{-1}$)                &              &2&75  &2.41\\
                            &H\,\RNum{2}   &1&142 &2.19\\
                            &              &2&88  &2.72\\
                            &PDR           &1&49  &2.2\\
                            &              &2&17  &2.36\\
\hline
\end{tabular}
\\Notes. Column 3 notes: (1)--non-outflow candidates, (2)--outflow candidates
\end{table*}

\section{SUMMARY}\label{sec5}

A search for outflows towards 694 star forming regions identified in previous studies based on the MALT90 survey has identified 188 high-mass outflow candidates, among which 85 clumps have SiO emission. The outflow properties were calculated for 105 sources with well-defined bipolar outflows and reliable distances. The parameters of these 105 sources may be underestimated to relate to temperature and opacity but overestimated due to include ambient material. Some factors have little effect on the overall distribution and correlations between individual quantities(e.g. \hcop abundance, and distance). The main results of this study can be summarized as follows:

\begin{enumerate}
 \item We identified 188 high-mass outflows from a sample of 694 clumps with a detection rate of approximately 27\%. We found that the outflow detection rate increases from the proto-stellar to the H\,\RNum{2} stage. A decrease in the detection rate at the PDR stage is likely a result of the outflow switching off during this stage.

 \item We found that there is an intimate relationship between outflow action and the presence of masers and that water masers may appear at a similar stage to 6.7 GHz methanol masers.

 \item Outflow action decreases infall detection rate at each evolutionary stage, and there is no obvious relationship between the infall mass rate and outflow mass rate of clumps.

 \item  The outflow action has a small contribution to the turbulence in the clumps, and the outflow contribution decreases with time as the outflow action ceases. Meanwhile, the outflow contribution to the turbulent energy is similar from the Proto-stellar to H\,\RNum{2} stages. Therefore, it can be concluded that there is no significant cumulative impact on the turbulence levels after repeated outflow action.

 \end{enumerate}

 Because MALT90 data have higher noise, some outflow candidates are likely not found and the reliability of some candidates may be low. This has some influence on our statistic results. Some data with lower noise is thus needed to further examine the accuracy of our conclusions. In addition, in order to exclude the influence of different clump masses. it is necessary to acquire some data for similar clump mass at the same stage to determine whether there is an obvious difference between clumps with and without outflow.

\section*{Acknowledgements}
This research has made use of the data products from the MALT90 survey, the SIMBAD data base, as operated at CDS, Strasbourg, France.
This work was funded by The National Natural Science foundation of China under grants 11433008, 11703073, 11703074 and
11603063, and The Program of the Light in China's Western Region (LCRW) under Grant Nos. 2016-QNXZ-B-22, 2016-QNXZ-B-23.
WAB has been supported by the High-end Foreign Experts grants Nos. 20176500001 and 20166500004 of the State Administration of Foreign Experts Affairs (SAFEA) of China and funded by the Chinese Academy of Sciences President's International Fellowship Initiative Grant No. 2019VMA0040.

\section{Appendix}
Table A1: the outflow properties of the blue and red lobes.\\
Table A2: The basic information of part clumps.\\
Figure B: Position-Velocity diagrams.\\
Figure C: The integrated intensity images of the blue and red wing.\\
Figure D: The integrated intensity contours of SiO emission.\\
\end{document}